\documentclass[useAMS,usenatbib]{mn2e}
\usepackage{graphicx,graphics,amsmath}
\usepackage{latexsym}

\usepackage{float}
\usepackage[caption = false]{subfig}

\begin{document}

\title{Combustion adiabat and the maximum mass of a quark star}

\author[Ritam Mallick \& Mohammad Irfan]
{Ritam Mallick$^{1}$\thanks{mallick@iiserb.ac.in}, Mohammad Irfan$^{1}$\thanks{mohai@iiserb.ac.in} \\
$^{1}$ Indian Institute of Science Education and Research Bhopal, Bhopal, India \\ }


\maketitle

\begin{abstract}
We solve the Combustion adiabat (CA) or the Chapman-Jouget adiabat equation to study the phase transition (PT) of a neutron star (NS) to a quark star (QS). 
The hadronic matter and quark matter equation of states are used to calculate the matter velocities on either side of the shock front.
The CA with the hadronic matter as an input is solved to obtain the corresponding quark matter values.
The maximum of the quark pressure is reflected in the retracing of the path in the CA curve. The downstream quark pressure maximum implies 
towards a maximum mass limit of a phase transformed QS which is different from the regular mass limit of an ordinary QS.
Further, the characterization of velocities suggest that the PT from NS to QS is not always feasible from the center of the star. 
The possible mode of combustion in NSs is likely to be a slow deflagration
in most of the low and intermediate density range.
The result is crucial and emphasizes on the fact that 
PT in NSs does not always starts from the center and sometimes a NS does not suffers a PT at all.
\end{abstract}



\section{Introduction}
Relativistic shocks are very common in astrophysical stellar objects and are associated with stellar winds, supernovae remnants, radio jets,
accretion onto compact objects and phase transition (PT) in neutron stars (NSs). The results of the relativistic Rankine-Hugoniot (RH) jump condition is well-studied \citep{landau} 
in connection with shocks, 
with their solution being still open for discussion. The RH condition can be used to derive a single equation relating the matter quantities across the discontinuity,  
known as Taub adiabat (TA) or shock adiabat relation \citep{taub}. Thorne \citep{thorne} showed that the TA method of solution could be carried over from non-relativistic 
to relativistic shocks.  In the TA, the initial and final states are the same functions of pressure, energy density and density (same EoS), and hence they lie on the same curve.
However, the form of the equation still holds if the initial and final states are not the same function of pressure, energy density, and density (different EoS). This happens 
because of the difference in the chemical energy of the initial and final state. Therefore, there is combustion from the initial state to the final state. Due to the difference 
in the EoSs the final state curve shifts from the initial state curve. The relation connecting the initial and final state is called the combustion adiabat (CA) or the 
Chapman-Jouget adiabat.
Several authors have used the relativistic RH condition to study various problems in astrophysics, but the study of CA in an astrophysical scenario is insufficient.

Relativistic shock phenomena have been widely discussed in recent years with regards to their connection to the PT in NSs.
NSs are thought to be the best candidates depicting a phase of strongly interacting deconfined matter. They serve as natural laboratories to study the low temperature and high density 
(or high baryon chemical potential) regime of the QCD phase diagram. Since the proposition that strange quark matter can be the stable configuration at such high densities 
\citep{itoh,bodmer, witten}, there has been considerable effort to investigate this possibility in astrophysical literature. One of the lines of investigation assumes it to be 
a simple first order PT from confined hadronic matter (HM) to deconfined quark matter (QM). Accretion process in NS can trigger this PT 
\citep{cheng2} or nucleation via seeding can also occur \citep{alcock}. 

The process of phase transition of matter at extreme astrophysical densities is very complicated and highly debated. Olinto \citep{olinto}
was one of the earlier physicists to study the combustion process. She viewed it as slow combustion process where an excess of down quark gets converted to strange quark via the 
weak interaction process.
Collins and Perry \citep{collins} argued that it is rather a two-step process, where the initial nuclear matter gets converted to final strange quark matter (3-flavour (3f)) 
via an intermediate 
step of 2-flavour (2f) quark matter. Horvath and Benvenuto \citep{horvath} argued that the convective instability could turn slow combustion (deflagration) process to a detonation. 
Cho et al. \citep{cho} used the hydrodynamic RH jump conditions to conclude that weak detonation is the correct mode of combustion. However, Tokareva et al. \citep{tokareva} 
suggested that the detonation modes can also be possible whereas Lugones et al. \citep{lugones} 
strongly advocated that the actual mode of combustion is strong detonation.

Bhattacharyya et al. \citep{mallick1} introduced two-step combustion involving the hydrodynamic jump equations to study the first step and slow weak combustion for the second
step. However, Drago et al. \citep{drago} argued the combustion mode to be a slow deflagration involving a mixed phase region. Recently Prasad \& Mallick \citep{mallick-apj}
did a dynamical evolution of the phase transition front and found that the PT takes not more than $50 \mu$s to occur.
Niebergal et al. \citep{neibergal} numerically solved for the front velocity involving the 
hydrodynamic equations along with neutrino emission and weak interaction of down to strange quark conversion and found that the combustion velocity of the weak process is very high.
Recently Furusawa et al. \citep{furusawa1,furusawa2} did a detailed discussion on scenarios involving shock induced and diffusion induced PT. 
In our analysis, we are dealing with shock-induced PT where the shock initiates a PT (or combustion) in NS. Instead of using the RH jump condition we employ the 
CA equation to study the PT of HM to QM in neutron stars. The CA is easily solvable as it a single equation devoid of any velocity terms.

The paper is arranged as follows: In section 2 we give the details of the EoS that we have used in our work. Section 3 deals with the study of the CA where 
we employ HM and QM 
EoS in the CA to show our results. Finally, in section 4 we summarize our findings and conclude from them.

\section{EoS and the M-R curve}

\begin{figure*}
\centering
\includegraphics[width = 3.14in,height=2.0in]{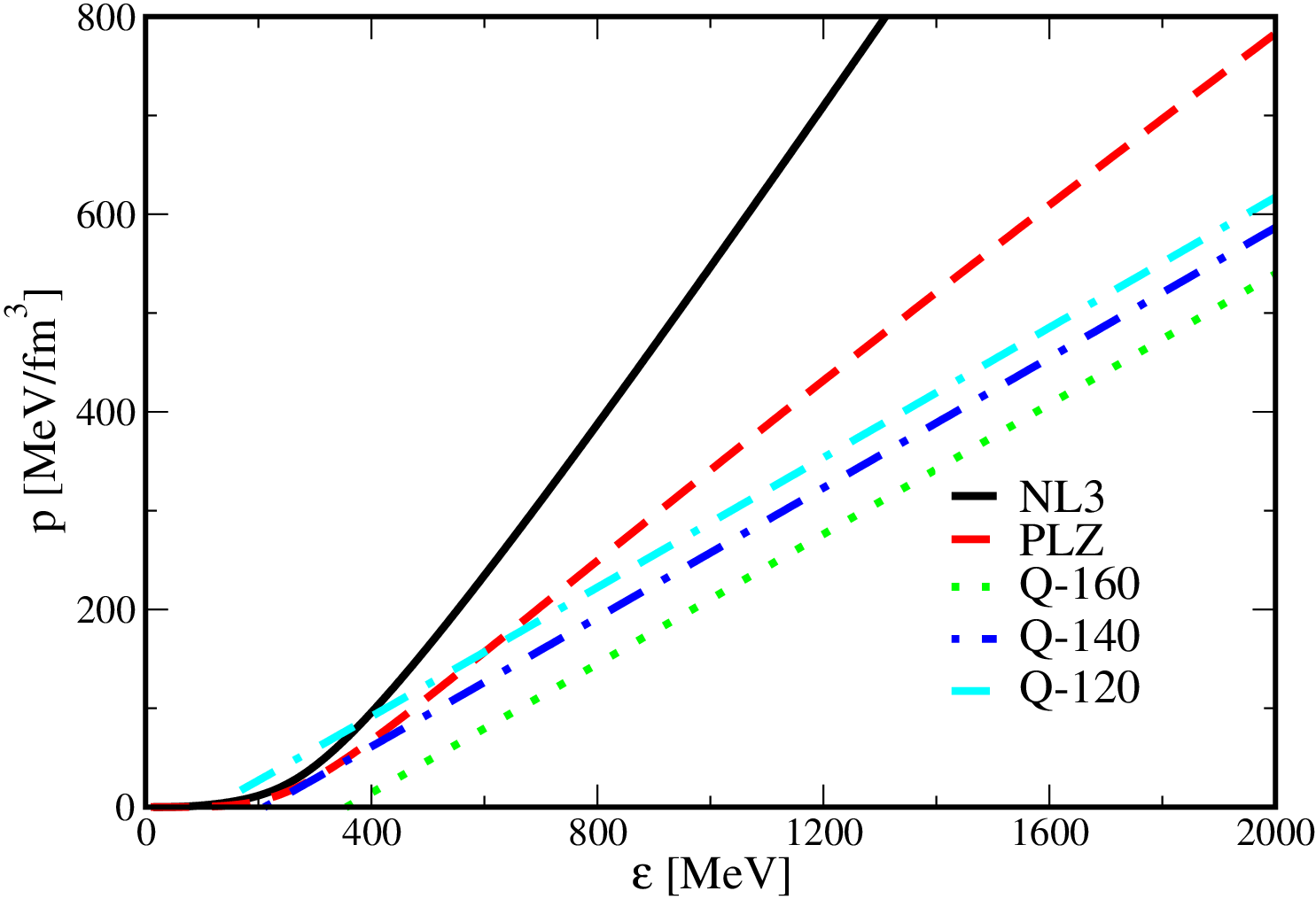}
\includegraphics[width = 3.14in,height=2.0in]{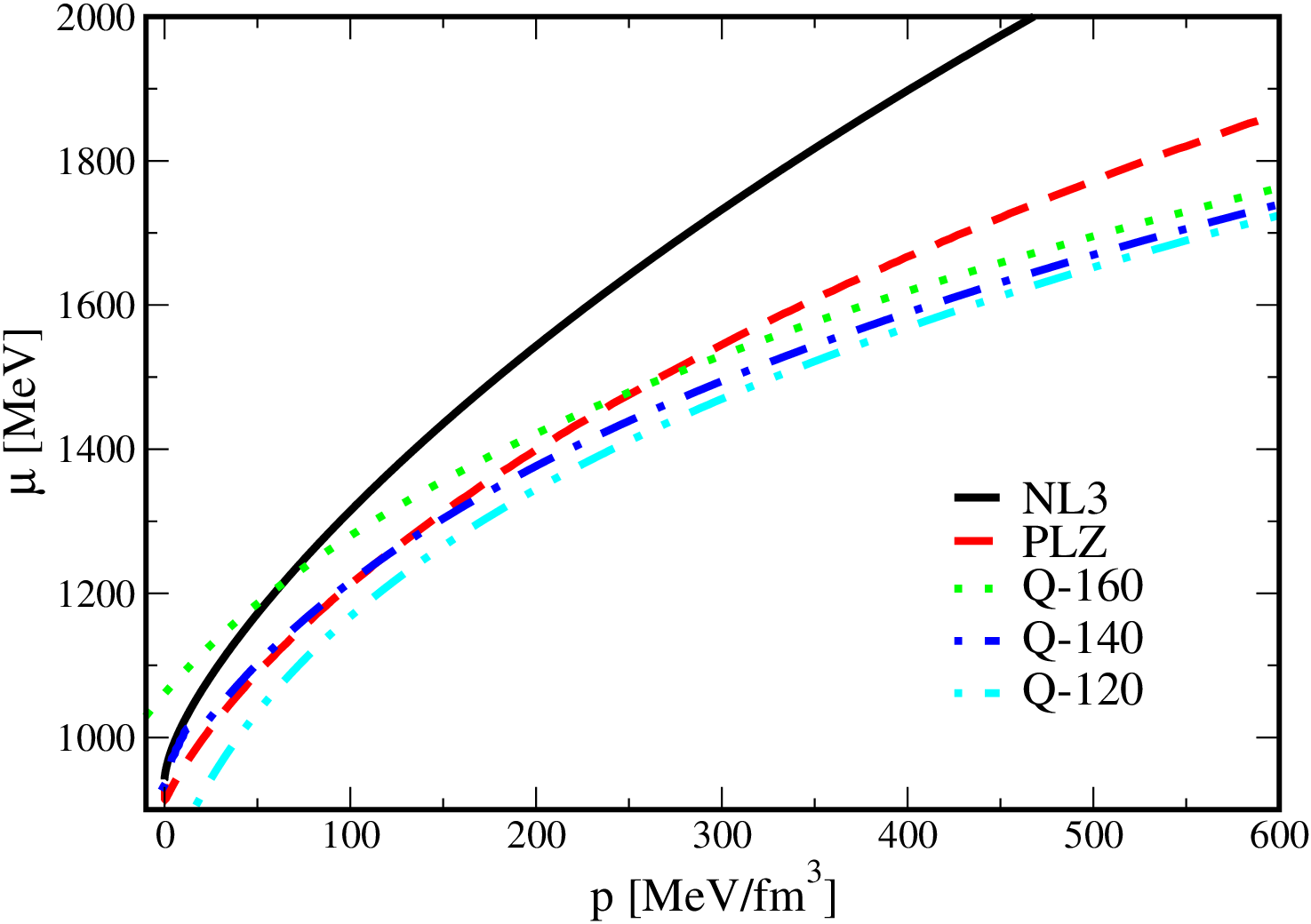}

\hspace{0.5cm} \scriptsize{(a)} \hspace{7.8cm} \scriptsize{(b)} 
\caption{(Color online) a) Pressures as a function of energy density ($\epsilon$) for HM (NL3 (black-solid) and PLZ (red-dash)) and QM (Q-160 (green-dot), Q-140 (blue-dash-dot) 
and Q-120 (cyan-double dash-dot)) EoSs are plotted. b) The chemical potential is shown as a function of pressure. While the terminology of the curves remains the same, the curve with a smaller chemical potential at a given pressure
is the more stable one. Though, the points, where the HM and QM curves coincide depicts the equilibrium PT from hadronic to quark matter.}
\label{fig-eosq}
\end{figure*}

We preferably use zero temperature EoS as we assume that the PT takes place due to density fluctuations in any ordinary cold pulsar. Although, the final burnt QM 
can have finite temperature depending on the EoS of matter on either side of the PT front.
However, in this calculation, we use zero temperature quark EoS. The finite temperature EoS is not likely to change our result drastically \citep{mallick-arx, mallick-apj,mallick-igor}.
For the hadronic phase, we adopt a relativistic mean-field approach which is generally used to describe
the NM in compact stars. The corresponding Lagrangian as given in the following form by \cite{serot,glen}($\hbar$=c=1)

 \begin{eqnarray} 
 {\cal L}_H = \sum_{n} \bar{\psi}_{n}\big[\gamma_{\mu}(i\partial^{\,\mu}  - g_{\omega n}\omega^{\,\mu} - 
\frac{1}{2} g_{\rho n}\vec \tau . \vec \rho^{\,\mu})- \\ \nonumber
\left( m_{n} - g_{\sigma n}\sigma \right)\big]\psi_{n} 
 + \frac{1}{2}({\partial_{\,\mu} \sigma \partial^{\,\mu} \sigma - m_{\sigma}^2 \sigma^2 } )- \frac{1}{3}b\sigma^{3}- \\ \nonumber
 \frac{1}{4}c\sigma^{4} - \frac{1}{4} \omega_{\mu \nu}\omega^{\,\mu \nu}+ 
\frac{1}{2} m_{\omega}^2 \omega_\mu \omega^{\,\mu} -\frac{1}{4} \vec \rho_{\mu \nu}.\vec \rho^{\,\mu \nu} + \\ \nonumber
\frac{1}{2} m_\rho^2 \vec \rho_{\mu}. \vec \rho^{\,\mu} 
+ \sum_{l} \bar{\psi}_{l}    [ i \gamma_{\mu}  \partial^{\,\mu}  - m_{l} ]\psi_{l}. 
\label{baryon-lag} 
\end{eqnarray}

The EoS contains only nucleons ($n$) and leptons ($l=e^{\pm},\mu^{\pm}$).
We assume the leptons to be non-interacting. Also, the nucleons interact with the scalar $\sigma$, isoscalar-vector $\omega_\mu$ and isovector-vector $\rho_\mu$ mesons respectively.
The adjustable parameters are fixed by matching the fundamental NM properties and that of the features of finite nuclei.
The hadronic EoS is modeled after the NL3 
and PLZ parameter setting \citep{serot,glen,reinhard} having only baryons. 

Three different quark EoSs (3f QM) are modeled after the MIT bag model \citep{chodos} having $u$, $d$ 
and $s$ quarks 
with bag pressures $B^{1/4}=160$ MeV (Q-160), $B^{1/4}=140$ MeV (Q-140) and $B^{1/4}=120$ MeV (Q-120) as plotted in fig \ref{fig-eosq}a. 
The corresponding masses of $u$, $d$ and $s$ quarks are $5$, $10$ and $100$ MeV respectively. For the quark interaction term (which can vary between $0$ and $1$) we have 
taken it to be $0.6$.
The EoS of the hadronic matter and quark matter are shown in fig \ref{fig-eosq}a. From here on QM would usually refers to 3f QM unless stated otherwise.
As evident from fig \ref{fig-eosq}a, for a 
fractional increase in energy density, the increase in pressure is the most for the NL3 curve (black solid line), hence implying it to be the hardest. Following
the same argument the Q-160 (green-dot) 
curve is the softest. The Q-120 curve is the stiffest among the quark curves, and intersects the two HM curves at low densities. 
The equilibrium PT from HM to QM occurs at densities where quark matter is the more stable form of matter. The relative stability of the HM to QM at different pressure 
(or densities) can be obtained from the $\mu$ vs. $p$ curve as shown in fig \ref{fig-eosq}b. For a phase to be relatively stable, the chemical potential ($\mu$) of the 
corresponding phase should be smaller than that of the other. 
From fig \ref{fig-eosq}b we find that as compared to the NL3 HM (black solid line), all the QM curves lie below it and are much stable at relative most pressures 
(or densities), apart from the point where the Q-160 (green dashed) curve cuts the NL3 (solid black line) and is indeed pivotal in determining the PT from hadronic 
to quark matter. Before this pressure value ($\sim 70$ MeV/fm$^3$) the Q-160 curve is seen lying above the HM curve for a small pressure range, ensuring that 
HM is the more stable form of matter in that region while beyond this point the Q-160 curve lies below the HM curve, rendering the latter 
to be relatively unstable than QM. For the PLZ HM EoS, although the Q-120 is always stable, the Q-140 and Q-160 EoSs 
attain stability at larger pressure values. The Q-160 curve cuts the PLZ curve at a pressure $\sim 260$ MeV/fm$^3$, whereas the Q-140 curve intersects it at a 
relatively smaller pressure value $\sim 110$ MeV/fm$^3$. Clearly, for the Q-160 EoS for QM there is a very limited window of pressure values where the QM can 
exist inside the star. Contrastingly, for Q-140 
EoS there is a considerable window of pressure values where QM can generate inside a star and proves to be aptly suited for our calculations. 
Henceforth, we will work with PLZ as our input HM EoS and Q-140 as our downstream EoS.

\begin{figure*}
\centering
\includegraphics[width = 3.14in,height=2.0in]{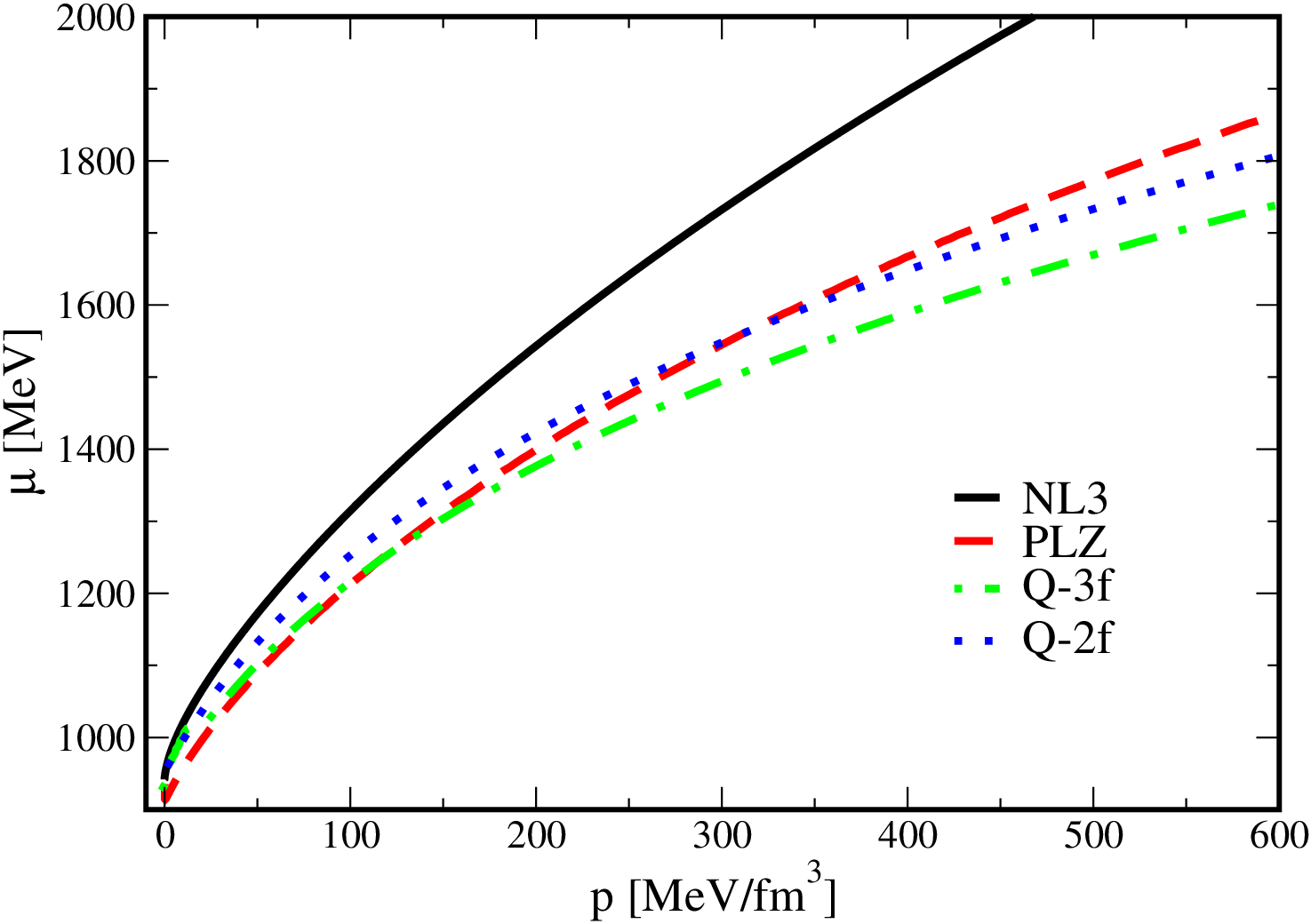} 
\caption{(Color online) The chemical potential is shown as a function of pressure for HM, 2f QM and 3f QM.  
The NL3 curve is marked by black-solid line, PLZ by red-dash line, 3f QM (Q-140) by green-dash-dot line and 2f QM (Q-140) by blue-dot line. 
The 2f curve is the metastable QM and the 3f curve is stable QM.
The points where the HM and QM curves coincide depicts the equilibrium PT from hadronic to quark matter (both for 2f and 3f).}
\label{fig-2f}
\end{figure*}

\begin{figure*}
\centering
\includegraphics[width = 3.14in,height=2.0in]{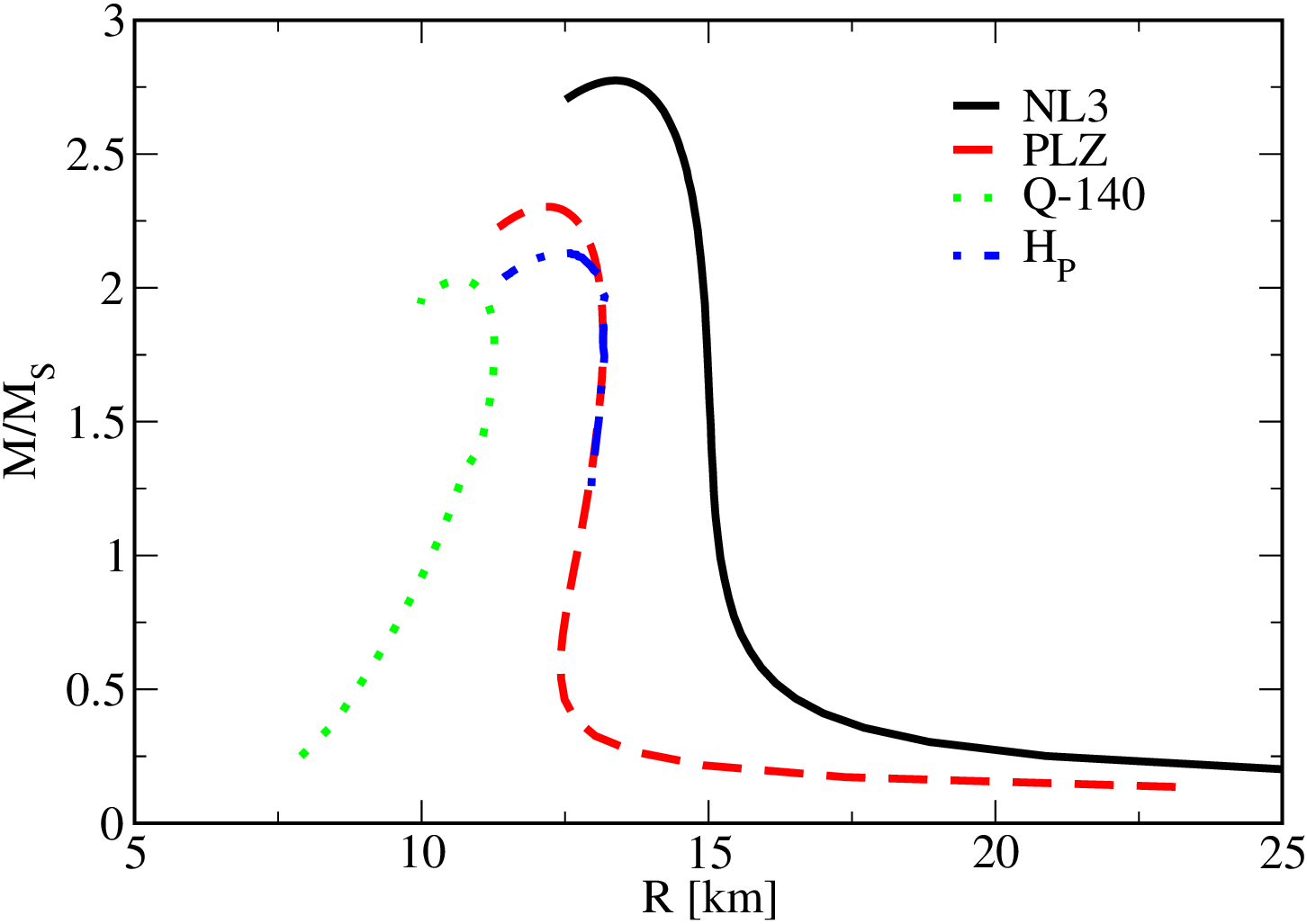}
\includegraphics[width = 3.14in,height=2.0in]{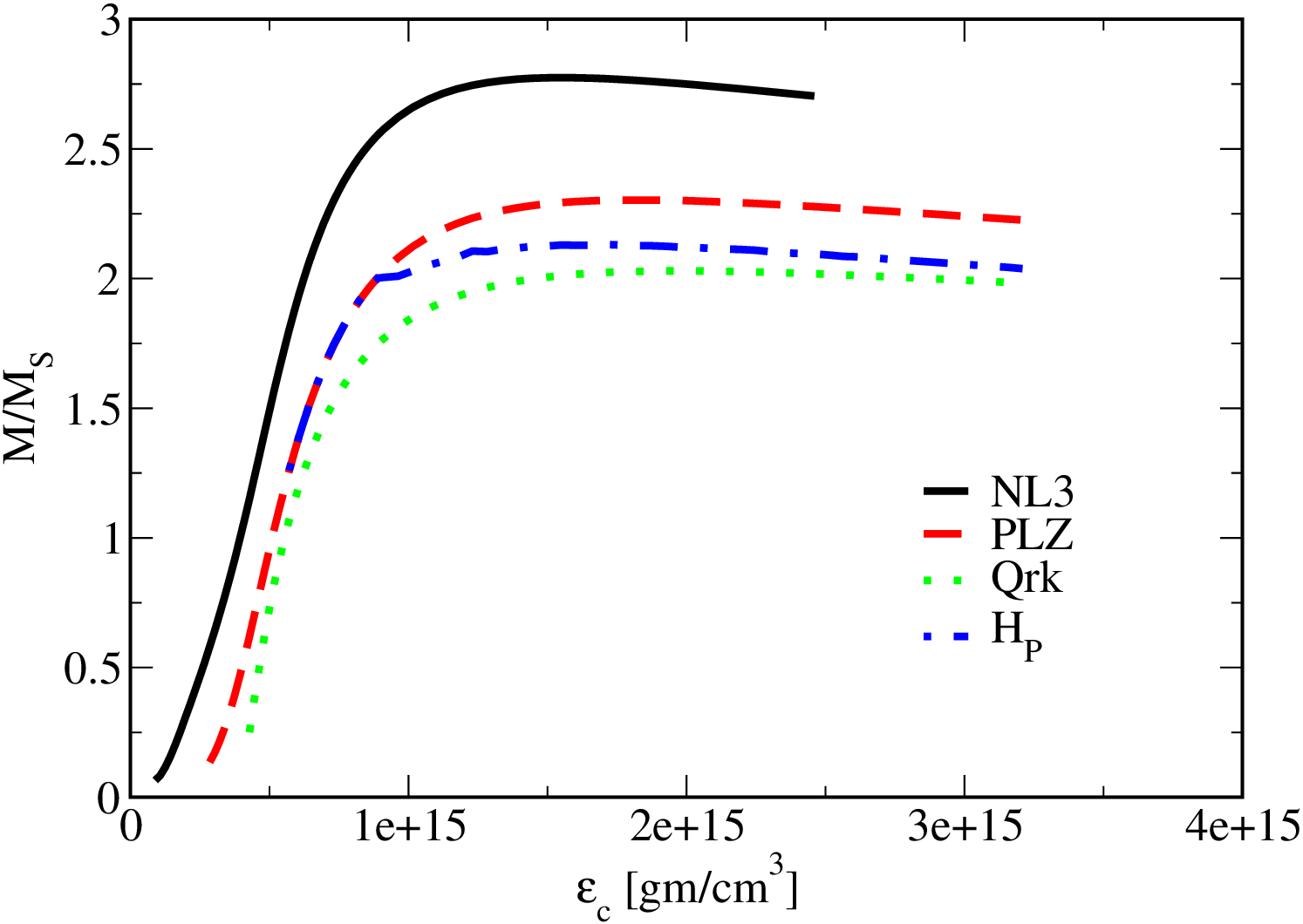}

\hspace{0.5cm} \scriptsize{(a)} \hspace{7.8cm} \scriptsize{(b)} 
\caption{(Color online) a) The mass-radius relations for pure hadronic matter corresponding to the NL3 (black-solid) and PLZ (red-dash) EoSs and for pure quark matter (green-dot) 
with $B^{1/4}=140$ MeV is plotted. The hybrid branch of PLZ and Q-140 is denoted by $H_P$ (blue-dash-double dot) for equilibrium PT. 
All curves are consistent with the recent limits of neutron star mass and radius. b) The mass as function of central energy density is depicted, 
while the nomenclature remains the same.}
\label{fig-mr}
\end{figure*}

From fig \ref{fig-2f} we see that the 3f QM (Q-140) is stable than the HM (PLZ) beyond pressure $p=110$ Mev/fm$^3$. Therefore, in the low-density region 
(corresponding to the pressure below $110$ Mev/fm$^3$) HM is the stable configuration. However, for the NL3 HM EoS, the curve for 3f QM (Q-140) always remains 
below The NL3 curve. Therefore, the 3-flavor QM is always
the stable state for such densities. This, tells us that for a PT taking place from PLZ NS to a quark star (QS) (with Q-140), the final QS is likely to be a hybrid star (HS). However, a PT 
from NL3 NS to Q-140 QS is more likely to produce a strange star (SS, stars comprising only of stable 3f QM). Fig \ref{fig-2f} also shows the metastable region of the 
2f QM. The PLZ and 2f curve (Q-140)
cuts at around pressure $p=340$ Mev/fm$^3$ and below those densities the 2f QM is metastable, and it has to finally go to 3f matter for absolute stability. 
Beyond those densities (corresponding to the pressure above $340$ Mev/fm$^3$) the 2f matter is stable than HM, and the star can spend a certain amount of time in this state, 
but ultimately it has to enter the
stable 3-flavor QM state. However, beyond $p=340$ Mev/fm$^3$ (corresponding to energy density $1.8\times 10^{15}$ 
gm/cc) we do not get any stable NS (fig \ref{fig-mr}). 

Using the given EoS for the HM and QM and solving the TOV equation \citep{tollman}, we can solve for the mass and radius of the star. Starting with different central densities we 
can also have the plot of the star sequence.
In fig \ref{fig-mr}a we have shown the mass-radius curve for the HM and QM EoS. The stiffest EoS generates the most massive stars which are evident from the fact that the 
star sequence
for NL3 EoS produces stars which can go up to 2.8 $M\odot$. However, recent studies show that NL3 is not so
consistent with the results from other NS bounds. Therefore, a more 
realistic EoS (PLZ parameter setting) is employed which produces stars as large as 2.3 $M\odot$. From now on our main results would be based on this PLZ hadronic matter EoS, 
however, for comparison, we would also show results related to NL3 EoS. 
The SS are obtained for Q-140 EoS (green dot).
Also the stable SS are above the given cut-off for mass. Our choice of the corresponding EoSs for HM and QM further generates HSs (for PLZ) 
which are also consistent with the mass bound. For a particular energy density, the HSs have lower masses than pure NSs (fig \ref{fig-mr}b). 
$H_P$ (blue dash-dot) denotes the hybrid branch corresponding to the PLZ EoS which emerges $\sim$ 12.5 km in fig \ref{fig-mr}a. 
The crossing between NL3 and Q-140 doesn't happen and therefore we do not get any HSs sequence for NL3 and Q-140.
The choice of EoS sets has enabled us to have considerable number of stars whose interiors are composed of quark core.
We also plot the mass against the central energy density of the star in fig \ref{fig-mr}b. Looking at the curve, we can identify the region of stable star sequence.
Starting from the point where the mass goes down with an increase in energy density is where the unstable NS sequences emerge. 

In our calculation, we assume that the phase transition from HM to QM starts at the center of the star due to some sudden density and pressure fluctuation. A shock wave is 
generated at the center of the star which propagates to the surface of the star. The shock wave has enough strength which inflicts a deconfinement of hadrons to quarks. 
This shock is accompanied with a combustion front which converts NM to QM. The QM at such high densities is more stable than ordinary NM; a QS is formed. 
If the shock wave propagates to the surface and expels the crust, 
it can even generate a stable SS. 
However, in a more general case, the shock wave stops at some point inside the star when the density becomes less, and the shock loses its strength. Thereby, we are left with a 
HS. A star with QM core and outer NM region. A QS is referred to those stars which have some amount of QM in them and include both SS and HS.

\section{Combustion adiabat}
The CA is derived from the RH jump conditions.
This relation is entirely different from the much-studied RH relation as it is only concerned with the matter
properties on either side of the front and does not involve any velocity terms. Once the CA is solved to obtain the final state, we find the velocities
from the matter properties.  The above analysis gives us insight into an entirely new physics. The results are quite robust and can be obtained using different sets of EoS. 

We start with the known RH condition which connects matter properties across the front. We assume that the shock front is at $y=z=0$ and is perpendicular to the $x$-axis.
Across the front, the two phases are related via the energy-momentum and baryon number conservation. They are given by 
\begin{eqnarray}
 w_h\gamma_h^2v_h=w_q\gamma_q^2v_q ;\\
 w_h\gamma_h^2v_h^2+p_h=w_q\gamma_q^2v_q^2+p_q ;\\
 n_hv_h\gamma_h=n_qv_q\gamma_q\equiv j ;
\end{eqnarray}
with, $w$ is the enthalpy ($w=\epsilon+p$), $u^{\mu}=(\gamma,\gamma v)$ is the 
normalized 4-velocity of the fluid and $\gamma$ is the Lorentz factor. We assume that the PT happens as a single discontinuity and the front propagates separating the two phases. 
Therefore we denote $'h'$ as the initial state ahead of the shock front; in our analysis, it is HM,
and $'q'$ as the final state behind the shock, the final burned QM.

The above equations can be used to obtain the CA where the baryon flux conservation equation (4) can be rewritten as 

\begin{equation}
 v_h\gamma_h = \frac{j}{n_h}, \hspace{2 em}  v_q\gamma_q = \frac{j}{n_q}.
\end{equation}
Substituting the above quantities in equation (3) we get

\begin{equation}
w_h\left( \frac{j}{n_h}\right) ^2 + p_h = w_q\left( \frac{j}{n_q}\right) ^2 + p_q.
\end{equation}
In terms of chemical potential, $\mu_i=w_i/n_i$, the above equation can be written as

\begin{equation}
p_q - p_h = j^2 \left( \frac{\mu_h}{n_h} - \frac{\mu_q}{n_q} \right) , 
\end{equation}

\begin{equation}
\Rightarrow -j^2 = \frac{p_q -p_h}{\mu_q/n_q - \mu_h/n_h} .
\label{cur}
\end{equation}

Using the definition of chemical potential and solving equations (2) and (4) we get the relation

\begin{equation}
\mu_h \gamma_h = \mu_q \gamma_q
\label{eq-mu}
\end{equation}

Now, we are in a better place to obtain our CA equation. Dividing eqn \ref{cur} by $(\mu_q/n_q + \mu_h/n_h)$, we get

\begin{equation}
j^2 \left( \frac{\mu_h ^2}{n_h ^2} - \frac{\mu_q ^2}{n_q ^2} \right) = \left( p_q - p_h \right) \left( \frac{\mu_h}{n_h} + \frac{\mu_q}{n_q} \right) .  
\end{equation} 

Substituting the value of $j=n_h u_h=n_q u_q$ (equation (4)) we obtain

\begin{equation}
(\mu_h u_h)^2 - (\mu_q u_q)^2 = (p_q - p_h) \left( \frac{\mu_h}{n_h} + \frac{\mu_q}{n_q} \right).
\end{equation}

Subtracting this from the square of the equation (\ref{eq-mu}), $(\mu_h \gamma_h)^2 - (\mu_h \gamma_h)^2 = 0$, while keeping in mind the definition of Lorentz factor and fluid 
4-velocity we obtain

\begin{equation}
 {\mu_q}^2 - {\mu_h}^2 = (p_q - p_h)(\frac{\mu_h}{n_h} + \frac{\mu_q}{n_q}).
\end{equation}

This is the required CA equation which can also be written in terms of a quantity $X$ defined as $X_i=w_i/n_i^2=\mu_i/n_i$ and takes the form
\begin{equation}
 w_qX_q - w_hX_h=(p_q - p_h)(X_h + X_q).
\end{equation}

The so-called downstream quantities (denoted by subscript ``q'') are calculated from the known upstream quantities (subscript ``h''). 
The equation for the Taub adiabat also has an identical form. However, there the upstream and downstream quantities lie on the same curve.
In this study, we use the CA equation with two different EoSs for the upstream and downstream quantities.
For the known EoS of the HM, we can plot a curve in the $X,p$ plane indicated by NL3/PLZ. For a particular density in the HM, we have the corresponding pressure and enthalpy
from the HM EoS. Using these as the input values for the CA equation we calculate the corresponding QM pressure, density, and enthalpy. Therefore, for a given initial state 
of the HM, the final burned downstream state will lie on a different curve corresponding to the QM, indicating a burning or in our case a PT. 

\begin{figure*}
\centering
\includegraphics[width = 3.14in,height=2.0in]{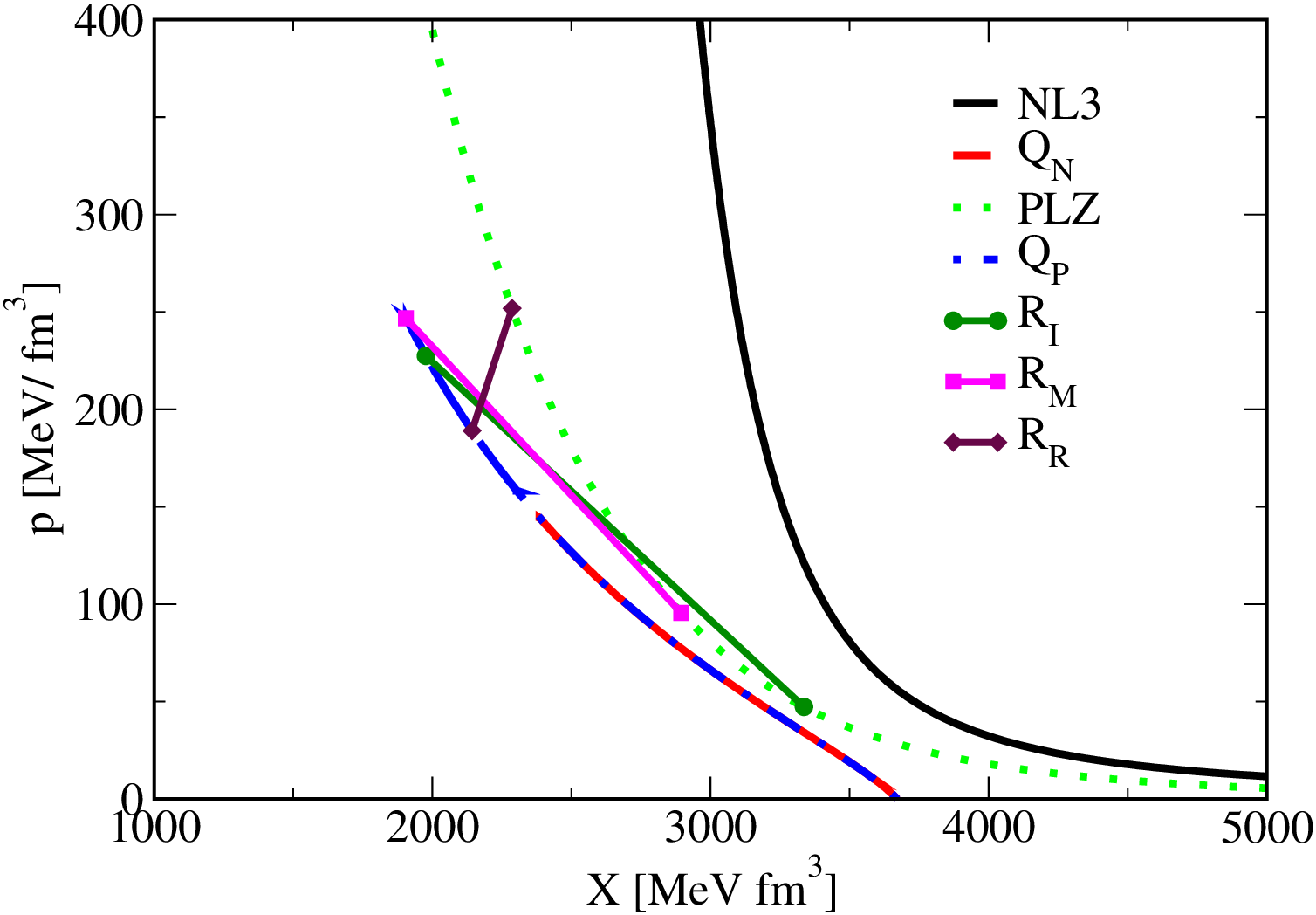}
\includegraphics[width = 3.14in,height=2.0in]{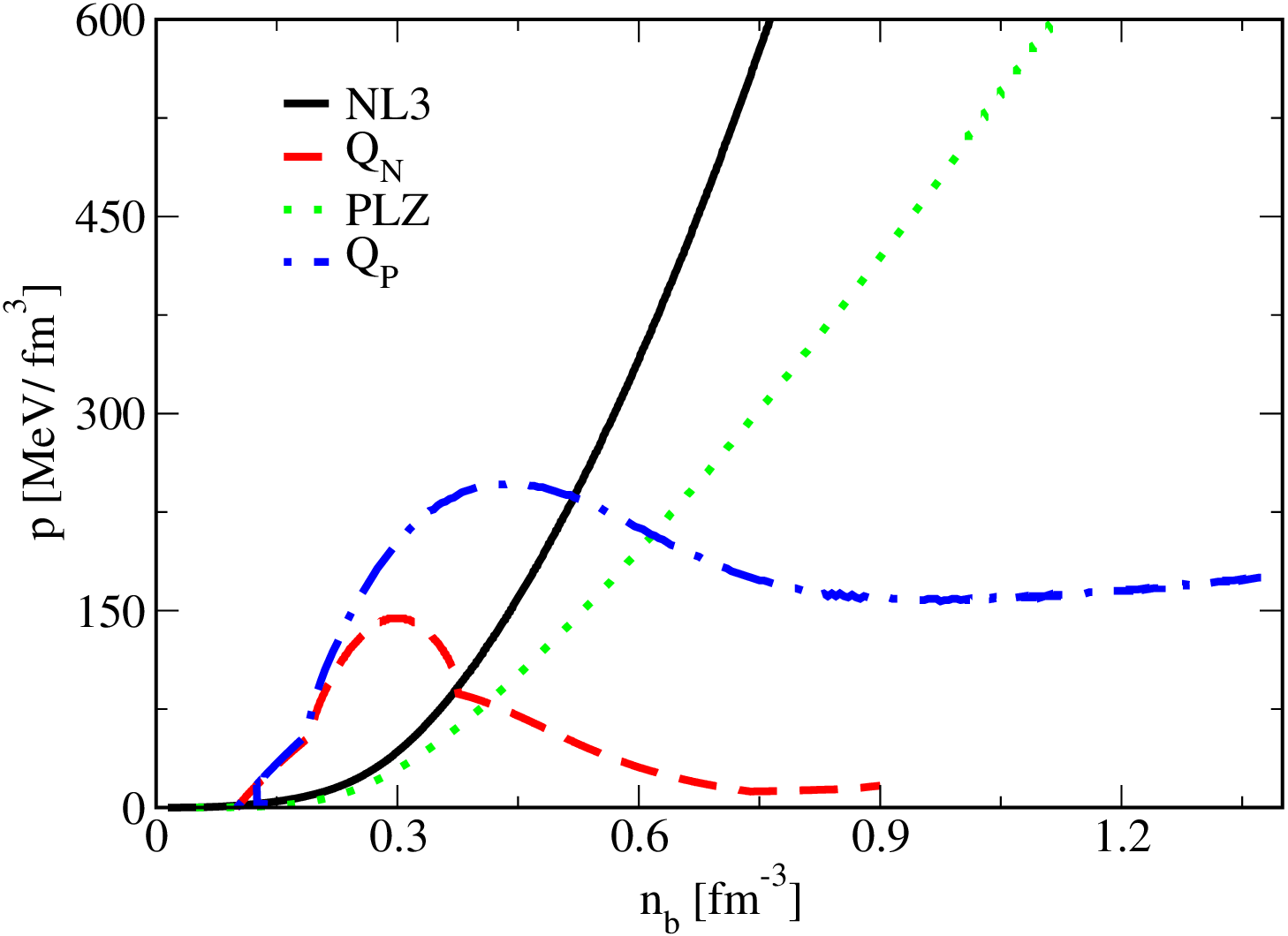}

\hspace{0.5cm} \scriptsize{(a)} \hspace{7.8cm} \scriptsize{(b)} 
\caption{(Color online) a) The CA ($p$ vs. $X$) curves for HM (NL3 (black-solid) and the PLZ (green-dot)) with their corresponding burnt states Q$_N$ (red-dash) and Q$_P$ (blue-dash-dot) also drawn. The QM EoS is assumed to be Q-140.
The quark curve of the PLZ (Q$_P$) extends much beyond the corresponding quark curve of the NL3 (Q$_N$).
$R_I$ indicates the regular jump from NL3 to its burnt state; $R_M$ indicates the maximum on the burnt QM state and $R_R$ marks retracing of the curve. b) $p$ as a function of baryon density 
($n_b$) for HM (NL3 and the PLZ) and their corresponding downstream QM (Q$_N$ and Q$_P$) curves are illustrated. The burnt pressure first rises and then decreases cutting their corresponding HM pressure curves at particular n$_b$ 
values.
(An animation of the CA is avaliable which is provided in the supplementary material ($anim.avi$). The animation shows the retracing nature of the CA which is discussed in the text).}
\label{taub1}
\end{figure*}

In the CA equation, while already having the EoS of HM, the upstream quantities ($\mu,p,n$) are known. Hence, solving the CA equation, we find the downstream 
pressure from which 
we calculate other quantities ($\mu,n$) using the known QM EoS. Therefore, we have one equation with only one unknown which is easily solvable. 
The slope of the “Rayleigh” line, connecting this initial point with the final point on the 
CA is proportional to $(\gamma_n v_n)^2$. 
Once the combustion adiabat equation is solved, we can calculate the velocity of the upstream and downstream quantities \citep{thorne}.

Starting from a point ($X_h,p_h$) (fig \ref{taub1}a) in the CA if we encounter some point ($X_h^/,p_h^/$) in the same curve, we have a shock wave. However, with the same starting point, if we reach a 
point ($X_q,p_q$) having different EoS (in this case Q-140), we will either have a detonation or a deflagration. 
Gradually changing the input values $(X_h,p_h)$ we generate the upstream curve. Taking the points of this curve as input and solving the CA for the QM EoS we generate the downstream curve ($X_q,p_q$). 
Initially, as the upstream curve goes up, the downstream curve also rises. The line marked with $R_I$ in fig \ref{taub1}a indicates a typical Rayleigh line of such a region. 

However, there is a maximum point on the burnt trajectory beyond which if we go northward along the HM 
curve the burnt trajectory goes down and retraces its path. The Rayleigh line connecting the maximum point of the burnt curve and the corresponding initial point of the upstream 
curve is marked with $R_M$. The retracing of the curve can also be aptly illustrated if we further draw a Rayleigh line connecting the upstream curve and the subsequently burnt curve 
(represented by $R_R$ in fig \ref{taub1}a). We find that the Rayleigh line now has an opposite slope and points downwards.
Such a nature portrayed by the curve, its maximum point and retracing of the track is quite robust and can be seen in hard as well as soft EoSs for both NL3 and PLZ curves. 

\begin{figure*}
\centering
\includegraphics[width = 3.14in,height=2.0in]{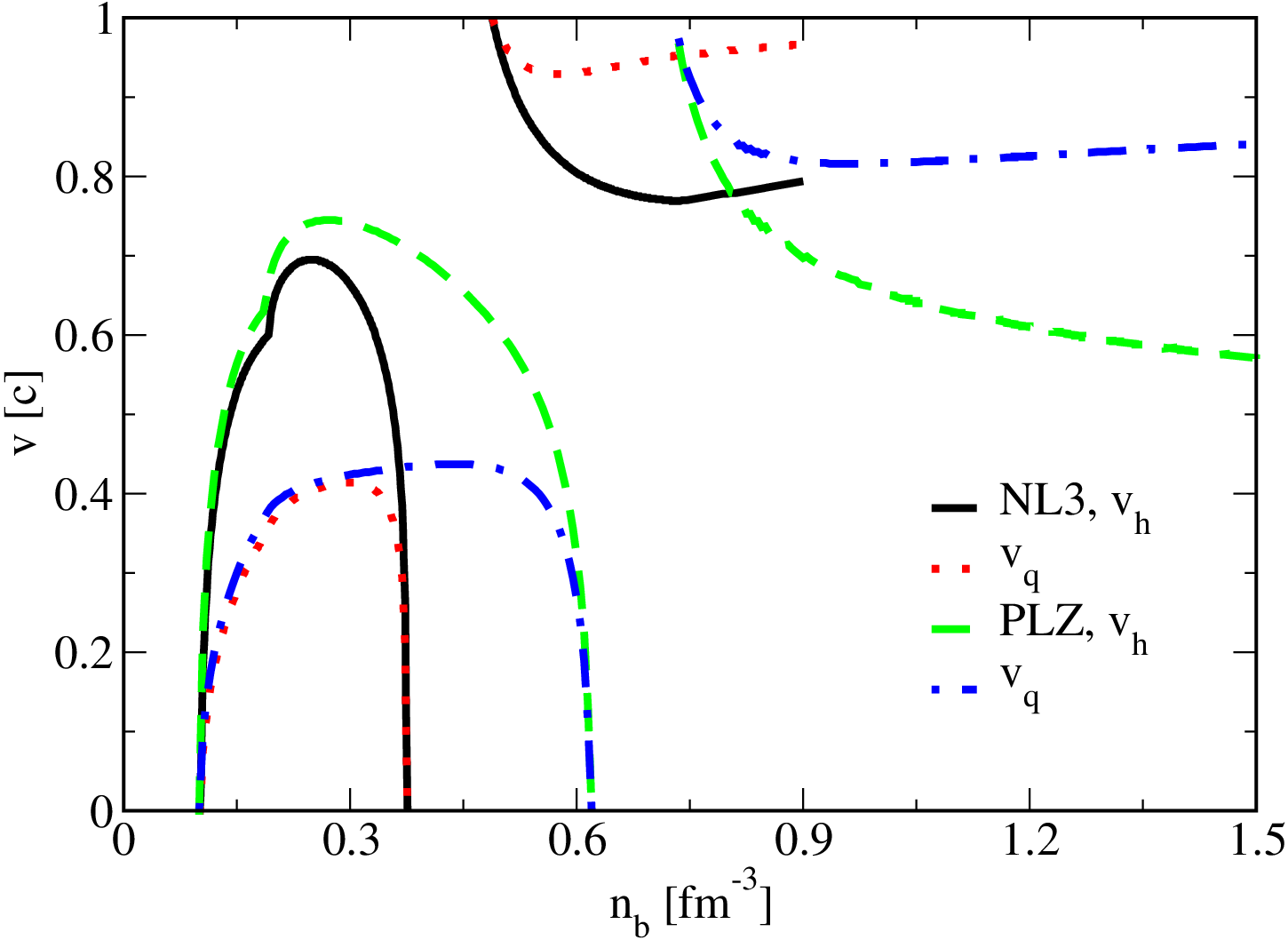} 
\caption{(Color online) The upstream ($v_h$) and downstream ($v_q$) velocities are shown as a function of $n_b$ for input HM EoS. Both $v_h$ and $v_q$ first increase till a 
point and then 
decreases and goes to zero. $v_h$ is always greater than $v_q$ at low density, however, the maximum point and the zero of the curves occur at same $n_b$ values. The velocities 
for NL3 EoS goes to zero at a much
smaller value of $n_b$ as compared to the PLZ velocities. The velocities again attain high non zero values at much higher densities (where $v_h$ is less than $v_q$).}
\label{fig-vel}
\end{figure*}

Fig \ref{taub1}b shows the plot of the upstream and downstream pressures as a function of the baryon number density ($n_b=n_h$). 
We find that the maximum of the calculated downstream pressure of fig \ref{taub1}b corresponds to the maximum point of the burnt curve beyond which it retraces its path. 
In fig \ref{taub1}b we get the point where the upstream and downstream 
pressures match playing quite a crucial role 
in determining the velocities of the two phases which we show later in fig \ref{fig-vel}.

From fig \ref{taub1}a and fig \ref{taub1}b, we can conclude that the pressure and therefore the energy and density of the burnt phase cannot increase beyond a certain 
point if we are 
considering a PT from state $h$ to state $q$. Therefore, if an NS with moderate central energy density suffers a PT to a QS, the central density of the QS would increase (quark EoS being more compact). 
This estimate is consistent with the initial part (lower pressure) of the CA curve where $p_q > p_h$. 
However, going to higher pressure values along the HM curve the corresponding pressure of the QM trajectory goes down from an absolute maximum point (marked as $R_M$ in fig \ref{taub1}a 
pink-square). After some density beyond the pressure maxima (the crossing point in fig \ref{taub1}b), the central energy density of the NS becomes larger than that of the QS.
Therefore, the resultant QS will be small. Such PT would release a substantially amount of energy which can have observational signatures like GW, 
neutrino beaming and gamma-ray bursts. On the other hand, it also confirms that the PT of an NS obeying such 
EoS can lead to a QS having an upper bound on the maximum mass. We note here that this constraint on mass is different from the Chandrasekhar mass limit on QS
\citep{banerjee} and is solely associated with QSs obtained as a consequence of the PT. 

The process of PT is through combustion. Previously, in literature there has been a lot of dispute whether the combustion would be a strong detonation or a slow deflagration. 
The combustion process can be deduced if we know the velocity of the respective phases. 
The velocities can be calculated from the upstream and 
downstream matter quantities \citep{thorne}. 
\begin{eqnarray}
 |v_h|=\left[\frac{(p_q-p_{ h}) (\varepsilon_q+p_h)}{(\varepsilon_q-\varepsilon_h) (\varepsilon_n+p_q)}\right]^{1/2}, \nonumber \\
~~|v_q|=\left[\frac{(p_q-p_{ h}) (\varepsilon_h+p_q)}{(\varepsilon_q-\varepsilon_h) (\varepsilon_q+p_{ h})}\right]^{1/2}.
\label{evel}
\end{eqnarray}

\begin{figure*}
\centering
\includegraphics[width = 3.14in,height=2.0in]{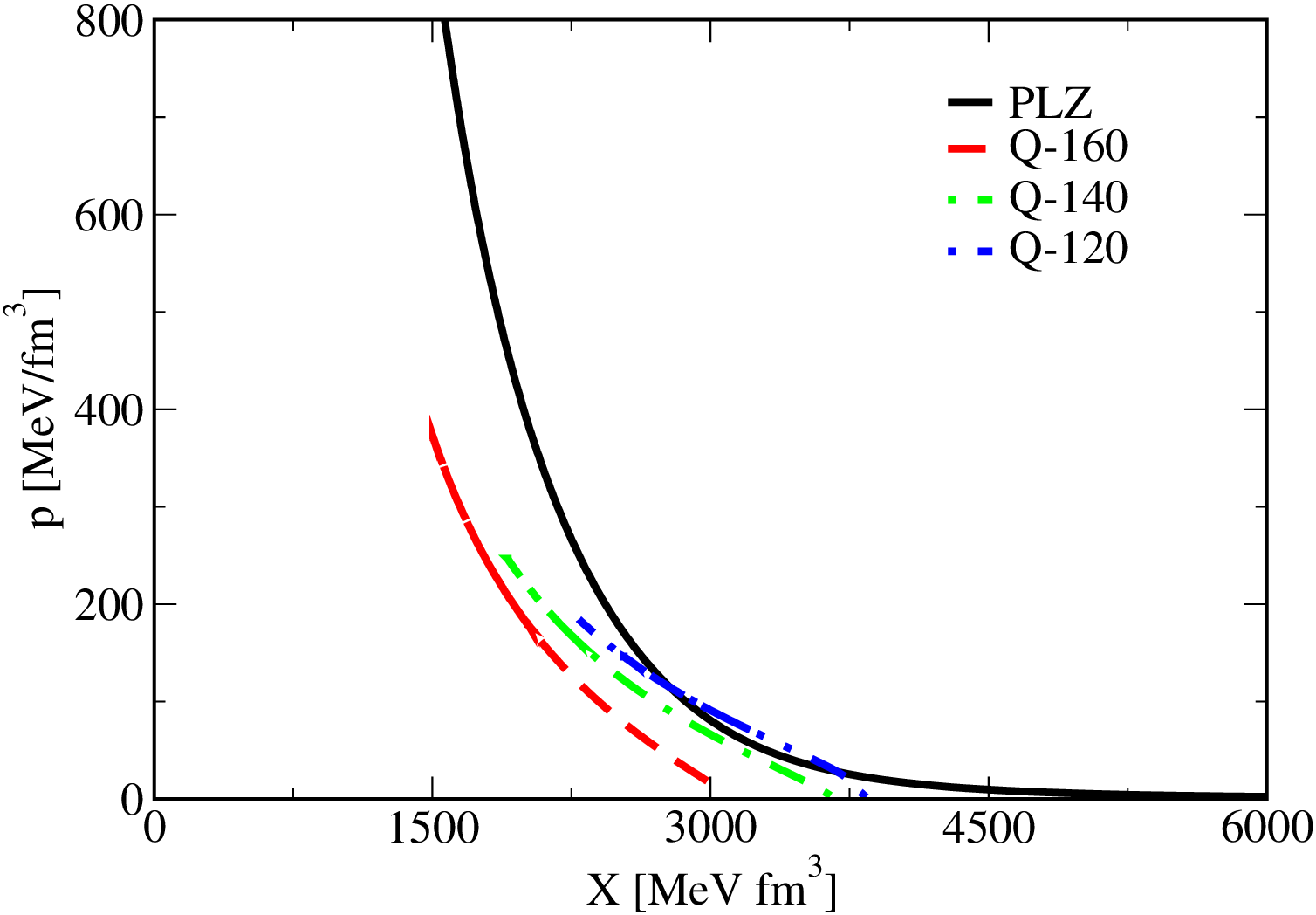}
\includegraphics[width = 3.14in,height=2.0in]{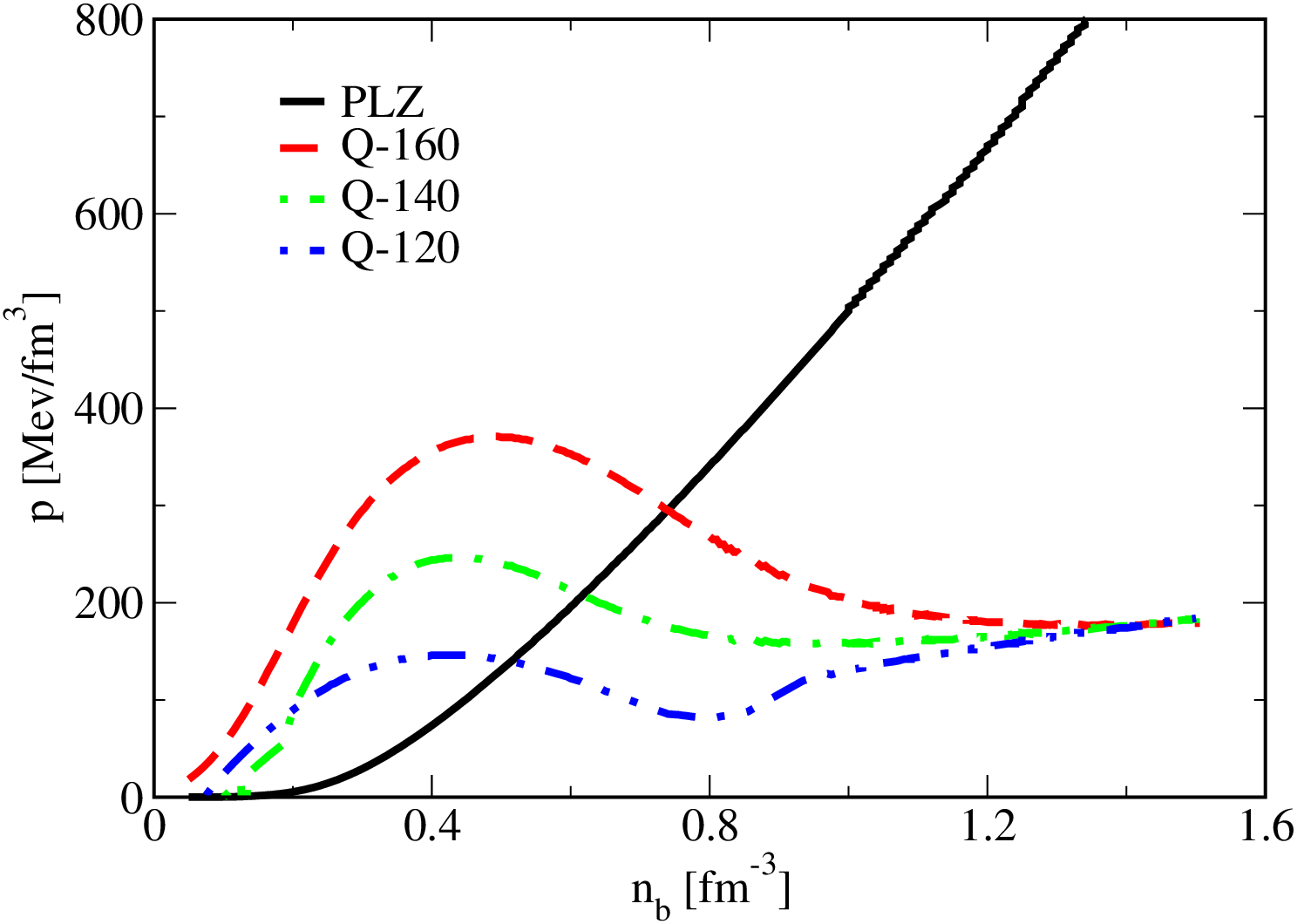}

\hspace{0.5cm} \scriptsize{(a)} \hspace{7.8cm}  \scriptsize{(b)}
\caption{(Color online) a) The CA curves for PLZ (black-solid) and the corresponding burnt states for three different quark EoSs with different bag pressure are drawn. b) p 
against $n_{b}$ curves for upstream HM and downstream QM corresponding to three different QM EoSs are illustrated.}
\label{fig-taq}
\end{figure*}

In fig \ref{fig-vel} we have plotted the upstream and downstream velocities as a function of baryon number density. 
Initially, the velocities (both $v_h$ and $v_q$) increase with an increase in density (increase in upstream pressure) and attains a maximum value. 
This maximum of velocity coincides with the point for which the downstream pressure is maximum and also with the maximum point in the CA curve. 
The speed then decreases gradually and becomes zero at the density where the upstream pressure equals the downstream pressure. 
Beyond those densities, the velocities are either zero or unphysical (more than the value of c). 
However, at much higher densities the velocities again become finite. 
At low densities the front velocity $v_f = \frac{v_h - v_q}{1-v_hv_q}$ is small, and the PT is slow. As density increases the burning becomes fast, however, it has an upper limit.
Beyond those densities, the burning again becomes slower and stops at much higher values. NSs with such densities (where velocity is either zero or unphysical) cannot undergo a PT to 
a QS. 

The variation of matter velocities and the comparison of the burnt and unburnt matter velocities is a valuable tool to understand whether a shock propagation is a detonation 
or a deflagration. If the speed of the burned matter is higher than unburnt matter, the PT is a detonation one, whereas if the velocity of the unburnt matter is higher than 
burned matter it is a deflagration.
Detonation is very fast burning whereas deflagration is slow combustion.
It can be classified as \\
a) $v_q > v_h$,  detonation. \\
b) $v_q < v_h$, deflagration. \\
Looking at the velocity curve, we can deduce the nature of combustion. In the low-density region where the velocities are finite and non zero, the combustion process is a deflagration or slow combustion.
Whereas, in the high-density range the combustion is a detonation. At high densities, a large fluctuation is needed to have velocities which are comparable to that of the velocity of light
and if this is attained the PT is a fast burning one. 
The maximum of the pressure curve (also the combustion adiabat curve) and with the help of fig \ref{fig-mr}b we can deduce the maximum bound of the QS formed as a result of PT. 

We assume that some density/pressure fluctuation at the center of the star starts the combustion process. For a low mass star (with low central density) the combustion leads to an
increase in pressure/density at the center of the star ($R_I$ has a negative slope (fig \ref{taub1}a)). Therefore, a less massive NS would result in a slightly more massive QS.
This process continues until the HM pressure reaches $205$ MeV $fm^3$ (the crossing point of the PLZ and $Q_P$ curve of fig \ref{taub1}b). 
If an NS with a central pressure below this value undergoes a PT, it would result in a QS (SS/HS) more massive than the initial NS. On the other hand, if an NS with central 
pressure greater than $205$ MeV $fm^3$ undergoes a PT, it would result in a QS less massive than the initial NS.

The maximum mass bound on the QS formed due to the PT of an NS can also be found using fig \ref{taub1} and fig \ref{fig-mr}.
The maximum of the quark pressure occurs at $p_Q=245 MeV/fm^3$ (fig \ref{taub1}a, b) and the corresponding pressure of the HM which this corresponds to is $p_H=91 MeV/fm^3$. 
This hadronic pressure corresponds to the energy density $\epsilon_H=8.1 \times 10^{14}$ gm/cc. Such 
central pressure of HM corresponds to a NS of $1.9 M_{\odot}$ (obtained from fig \ref{fig-mr}b). The maximum of the quark pressure ($p_Q=245 MeV/fm^3$) corresponds to energy
density of $\epsilon_Q= 1.7 \times 10^{15}$ gm/cc. 
From our discussion in section II, we see that the PT from NS (for PLZ EoS) to QS (Q-140) would probably give rise to an HS instead of an SS. This is because at low densities PLZ 
HM EoS is more stable than 3f QM EoS. Therefore, in a star with central energy density $\epsilon= 1.7 \times 10^{15}$ gm/cc, at the core, the 3-f QM is the stable ground state. 
However, as we go outward towards the surface of the star the density (therefore pressure) decreases. And at a certain radial point inside the star close to the surface
of the star the HM again becomes the stable state (corresponding to the crossing point of fig \ref{fig-eosq}b). Therefore, the PT or the combustion would stop at that radial point.
And thereby we have an HS with a quark core and hadronic outer surface. The corresponding mass of the HS is about $2.1 M_{\odot}$.
For such a star, the radial point where the combustion process is likely to stop is when the pressure at that point reaches $91 MeV/fm^3$. This happens at about a 
distance of $6.7$ km from the center of the star. Therefore, we have a star comprising of a quark core and outer nuclear matter, giving rise to a HS.
Furthermore, we should mention that the stopping of the combustion process may depend on several other factors which are solely obtained by dynamic calculation.
It may so happen that the PT propagates to the surface of the star, then we would get an SS. 
This corresponds to a SS of $2 M_{\odot}$. The maximum mass of the QS (HS/SS) formed after the PT almost coincides with the maximum mass of the QS obtained from the EoS 
curve for QS (fig \ref{fig-mr}a). However, the bound on the maximum mass for which the PT can take place for NS is much lower than the upper bound on the mass obtained from the 
PLZ EoS (the former being $1.9 M_{\odot}$ and the latter being $2.3 M_{\odot}$). For such QSs formed as an aftermath of PT; the combustion process is most likely to proceed 
as slow combustion. 

\begin{figure*}
\centering
\includegraphics[width = 3.14in,height=2.0in]{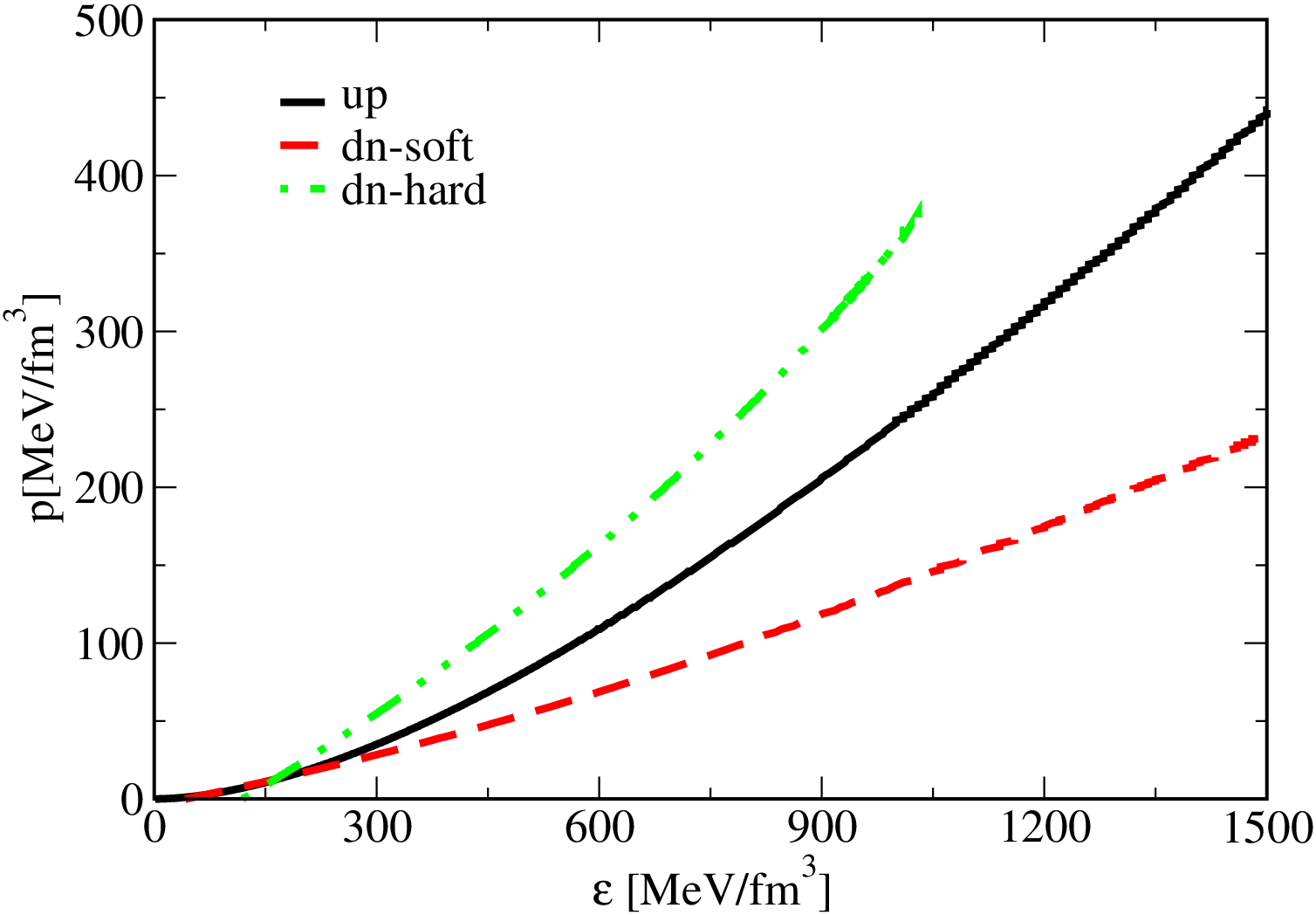}
\caption{(Color online) p as a function of $\epsilon$ plots are shown for the polytropes EoS. The polytrope describing the upstream (black-solid) lies in between the 
soft downstream (red-dash) and hard downstream (green-dash-dot) polytropes.}
\label{fig-eospoly}
\end{figure*}

The velocity curve (fig \ref{fig-vel}) gives the range for which the PT is possible (the non-zero values of velocities). For the PLZ EoS the range is from $n_b = 0.1 - 0.62 $ fm$^{-3}$.
However, moving from the lower densities towards the point where the hadronic and quark pressure coincides, the mass of the final QS is greater than the mass of initial NS. 
This can only be possible if some other form of energy is being converted to mass-energy, which can either rise due to a sudden slowing of the NS or sudden cooling of a hot 
proto-neutron star. 
The PT in this density range is most likely to happen via slow deflagration.

In the high-density range beyond $n_b =0.72$ fm$^{-3}$ the PT can also happen via strong detonation. In this region, the mass of the final QS is less than the mass of the initial NS. 
Therefore, for the shock-induced PT from PLZ NM to Q-140 QM, the central density range or the core density of NS where this
PT is more likely lies in the range between $n_b = 0.1 - 0.62 $ fm$^{-3}$ and again for $n_b  > 0.72$ fm$^{-3}$. 

For stars with central densities lying between $0.62-0.72$ fm$^{-3}$, the PT or the combustion front cannot start at the center of the star. However, at some radial point in 
those stars, where the central density is less than the $0.62$ fm$^{-3}$ some density fluctuation can give rise to a combustion front. The combustion front can propel to 
lower densities. However, the exact dynamics of the propagation of the detonation front to the core needs more detailed dynamic calculation. This is an interesting case, however, in
the present context we cannot precisely comment on that and needs a much more detailed analysis.

The mass limit and the density range of PT obtained from this calculation depends on the EoS sets we choose. For the NL3 and Q-140 EoS, the combustion is more likely to produce an SS.
This is because the 3-f QM (Q-140) is always stable than the NL3 EoS (the $\mu$ vs. p curve for QM always remains below the NL3 curve, fig \ref{fig-eosq}b). The maximum bound on 
the NS for which the PT 
can occur is $1.74$ solar mass, and the resultant mass of the SS bound is $1.77$ solar mass. 
Both the mass bounds are far lower than the maximum bound on the mass limits of their respective EoS. The final SS have a mass greater than the initial NS 
in the central density range $n_b = 0.1 - 0.375 $ fm$^{-3}$. The mode of combustion in this density range is slow combustion.
However, in the density range, $n_b  > 0.49$ fm$^{-3}$ the final SS is less massive than the initial NS, and the PT occurs via detonation. 
In the high-density range after a specific density ($n_b  = 0.65$ fm$^{-3}$), the initial NS becomes unstable as obtained from the mass-energy graph. However, in such stars, PT can occur to result in QS with 
lower mass and with a stable configuration.
For stars with central density in the range between $n_b = 0.375 - 0.49 $ fm$^{-3}$, the PT cannot start from the center.

To generalize our calculations, the HM (PLZ) EoS curve is plotted against three different quark EoSs having different bag pressures. The $\epsilon$-p plot in fig \ref{fig-eosq} 
portrays the HM (PLZ) curve along with a few quark EoSs employed for our calculation (Q-120, Q-140, and Q-160).
The CA curve (fig \ref{fig-taq}a) and the $p-n_{h}$ plot (fig \ref{fig-taq}b) for the aforementioned three different quark EoSs are plotted. As found earlier, the retracing 
nature is depicted in the CA plot in fig \ref{fig-taq}a. Given the dependence of $p$ on $X$ (fig \ref{fig-taq}a), we find that there occurs a downstream pressure maximum 
(fig \ref{fig-taq}b) for all the three cases which aptly portrays this retracing nature making our findings more profound. 
From fig \ref{fig-taq}a we find that as the EoS becomes stiffer, the CA curves shifts to the right. The maximum pressure from which the curve starts to retrace also decreases 
with an increase in the slope of the downstream EoS. Similar conclusion is also evident from fig  \ref{fig-taq}b.

Since the Q-120 curve (fig \ref{fig-taq}a) cuts the upstream HM (PLZ) curve, we therefore get a dip in pressure in the corresponding $p-n_{h}$ plot blue dash-dot in fig \ref{fig-taq}b) 
at densities $\sim$ 0.8 fm$^{-3}$, whereas at subsequent higher densities all the three curves tend to merge attaining the same pressure values.   
 
\begin{figure*}
\centering
\includegraphics[width = 3.14in,height=2.0in]{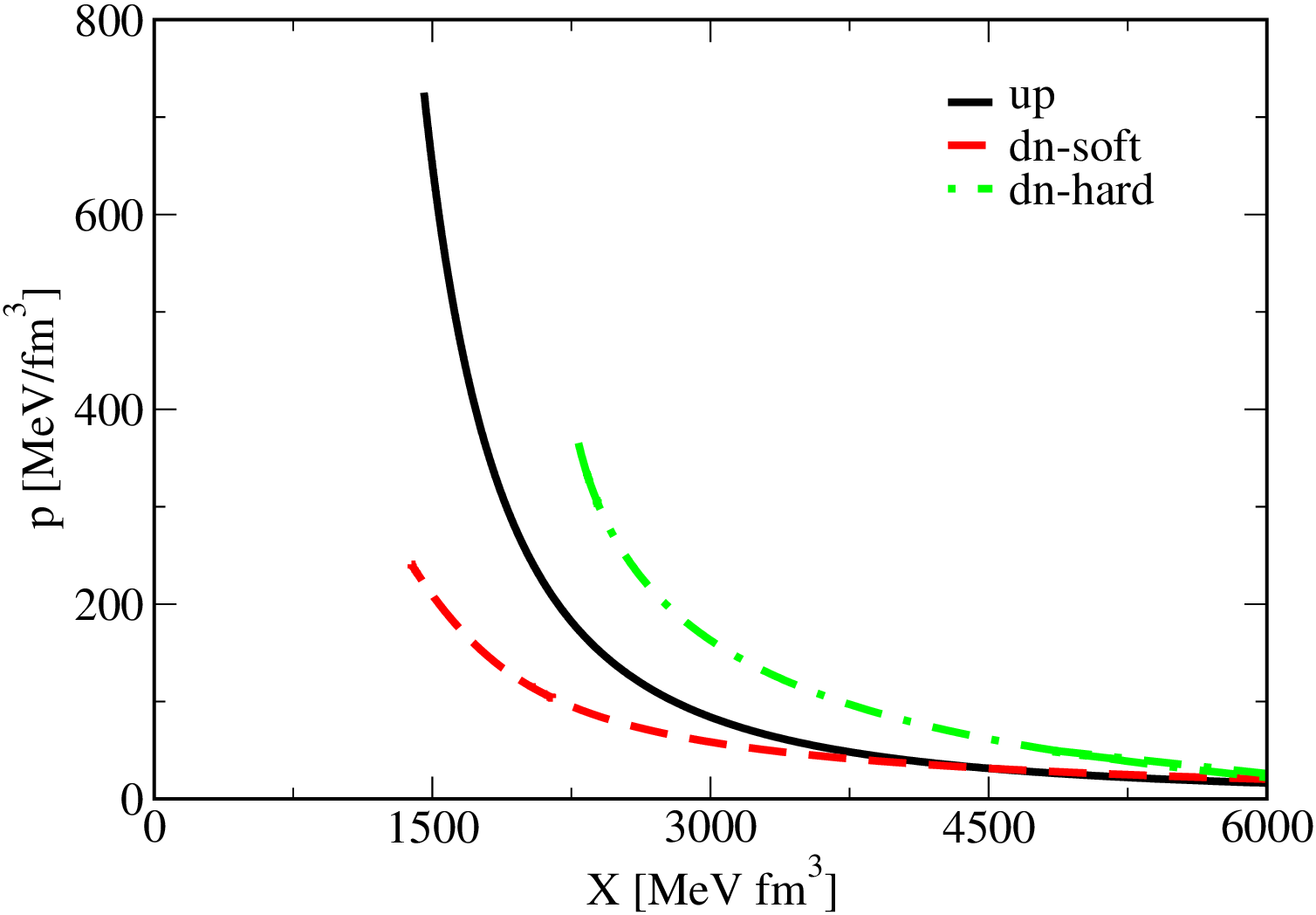}
\includegraphics[width = 3.14in,height=2.0in]{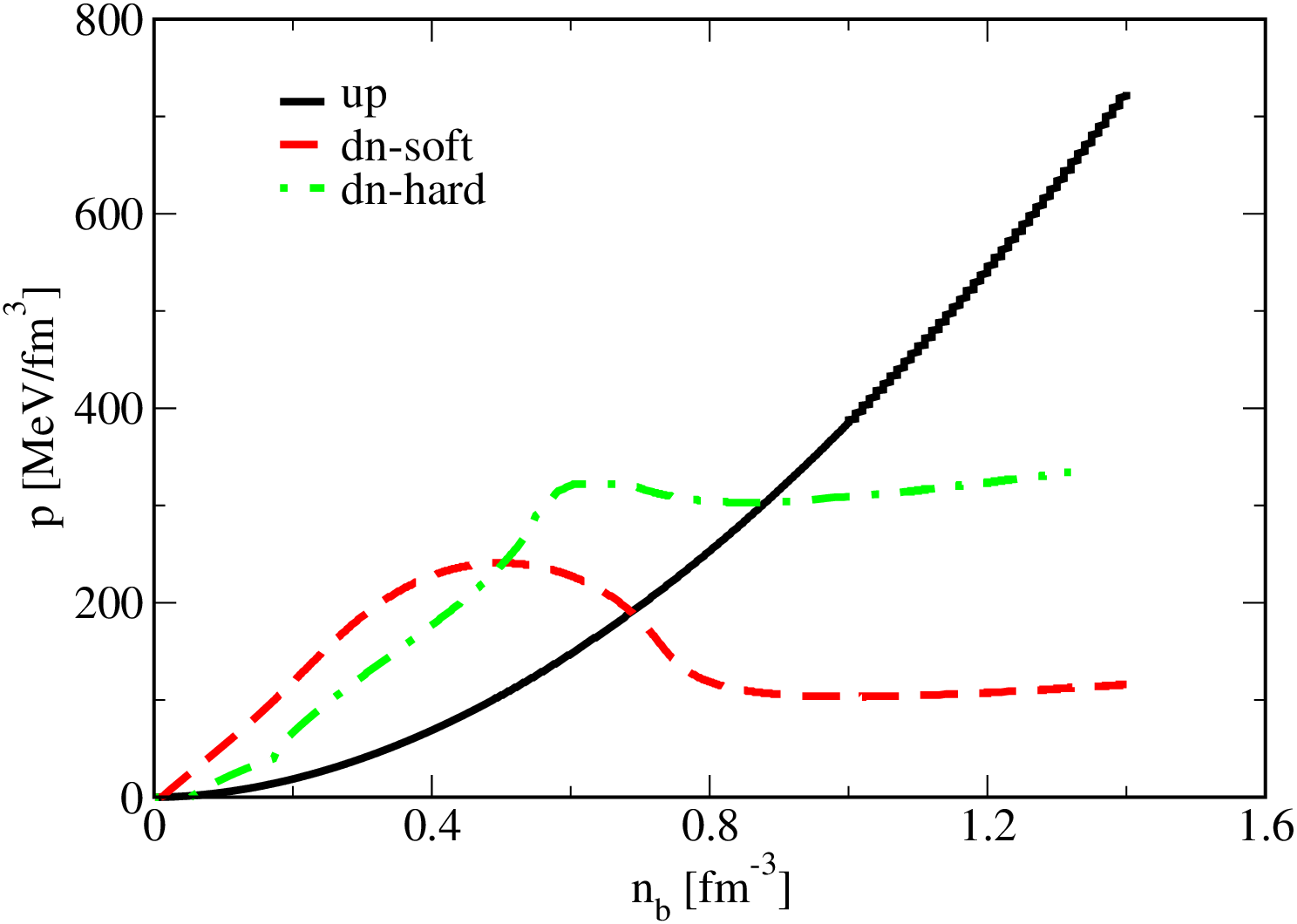}

\hspace{0.5cm} \scriptsize{(a)} \hspace{7.8cm}  \scriptsize{(b)}
\caption{(Color online) a) The CA curves corresponding to the polytropes for both HM (black-solid) and the QM (soft (red-dash) \& hard (green-dash-dot)) EoSs is plotted. b) 
p against $n_{b}$ curves for the upstream polytrope is marked by up (black-solid) and the downstream soft polytrope by dn-soft (red-dash) \& downstream hard polytrope by dn-hard
(green-dash-dot) is plotted.}
\label{fig-tapoly}
\end{figure*}

NSs can also be modeled by the polytropic EoSs. It is often used to construct the profile of stellar objects and simply follows a power law dependence of degeneracy pressure 
with the mass density i.e. $P=K\rho^{\gamma}$. However, to replicate a profile for real EoS (obtained by Relativistic Mean Field calculations) one needs both energy density and 
number density as a function of pressure. Eventually, 
one of the ways to construct a polytropic EoS was given by \citep{bonazzola}, where he described the polytrope as 
\begin{equation}
\epsilon (n) = m_{B}n+\frac{k\epsilon_{0}}{\gamma-1} \left( \frac{n}{n_{0}}\right) ^{\gamma}  \label{epoly}
\end{equation}
\begin{equation}
p(n) = k\epsilon_{0} \left( \frac{n}{n_{0}}\right) ^{\gamma} ,\label{ppoly}
\end{equation}
with n being the baryon number density, $\epsilon$ the energy density, p the pressure, k and $\gamma$ the dimensionless parameters, m$_{B}=931.2 $MeV (baryon mass) 
while n$_{0}$ and $\epsilon_{0}$ ($=m_{B}n_{0}$MeV fm$^{3}$) are number density and energy density at saturation.

While considering the polytropes, we adopt models for the QM EoS such that it lies on either side of the HM curve as depicted in fig \ref{fig-eospoly}. 
We have assumed our parameters (k, $\gamma$ $\&$ n$_{0}$) in such a way that one of the burnt curves lie above the HM plot while the other curve lies below it 
(fig \ref{fig-eospoly}). 
However, in astrophysical scenarios, it is generally found that the QM curve usually lies below the HM curve in the p-$\epsilon$ plot (fig \ref{fig-eospoly}). 
For our calculations, the polytrope for the upstream HM is constructed substituting k=0.038, $\gamma$=1.88, n$_{0}$=0.135 in eqns \ref{epoly} and \ref{ppoly}, 
while k=0.08, $\gamma$=1.5, n$_{0}$=0.14 make up the downstream soft QM EoS and k=0.054, $\gamma$=2.25, n$_{0}$=0.122 comprise the downstream hard QM EoS.  

Fig \ref{fig-tapoly}a shows the CA curves and fig \ref{fig-tapoly}b depicts the pressure curves for the polytropic EoS. As usual, the retracing nature is observed in the CA figure with
the hard downstream curve lying on the right of the HM curve and soft downstream curve on the left.
A rather intriguing nature of the pressure curve (fig \ref{fig-tapoly}b) is observed corresponding to the downstream hard QM EoS, i.e., which lies above the upstream 
HM curve in fig \ref{fig-eospoly} and to the right of HM curve in fig \ref{fig-tapoly}a.  The pressure initially increases uniformly, but the maxima are sharper and occur at a 
larger density corresponding to the uppermost point in the CA curve and saturating at a much higher pressure value.
From the CA curves and the maximum pressure curves plotted for realistic EoS and polytropic EoSs, one can safely conclude that the maximum and the retracing nature of the CA 
is quite a robust phenomena. 

\section{Summary and Conclusion}
To summarize, we have employed CA as a tool for studying the PT from NS to QS. Studies using the RH equations have been done earlier, where three equations are solved for three
unknowns. The velocities occur as an additional parameter for these equations. However, this CA technique is independent of velocity, and only the upstream and downstream EoSs 
are enough to study their PT. Our main results are shown using two HM EoS (NL3 and PLZ parameter setting) and a QM EoS 
(with bag pressure ($140$ MeV)$^4$). For comparison, we have also shown results with a single HM EoS (PLZ) and for three QM EoSs having three different bag pressures
(Q-120, Q-140, and Q-160). EoS having polytropic form has also been used to check our results.

Using HM EoS for the upstream curve and the QM EoS for the downstream burnt curve we solve the CA to obtain the respective downstream quantities. The whole of the downstream 
CA curve can be generated by varying the initial input upstream parameters. Solving the CA leads to a maximum pressure in the downstream QM state. After the maximum, the CA 
retraces its path. The retracing nature of the burnt QM CA curve is quite robust and is obtained for both hard and soft EoSs.  Although the upstream HM pressure always increases, 
the burnt QM pressure has a maximum. 
The retracing nature of the CA arises due to the occurrence of the maximum pressure of the burnt QM. 
This nature is not only limited to EoS of HM or QM but can also be extended to polytropic EoS which is sometimes used to calculate NS properties.
The above analysis of the CA indicates at a maximum limit on the QS (HS/SS). This mass limit is smaller than the Chandrasekhar mass limit for QSs. 
The study of the velocity to characterize the 
PT shows that the PT of NS to QS does not always start at the center of the star. For some particular central densities, the PT cannot trigger from the core, and the combustion begins at some radial
point in the star located at some distance from the center.

Observing the criterion for detonation and deflagration from the velocity of the respective phases, we found that the velocity of the NM in the low and intermediate density 
range is higher than that of the QM, which is the condition for a deflagration. For very high densities there is a possibility of detonation. Also, there is a 
considerable density range for which velocity is either zero or becomes unphysical. We can hence conclude that for these particular densities at the star center the PT 
process cannot commence from the star center.
We can also conclude that the burning process at the star center most likely starts as a deflagration process. 
For more massive stars the detonation process is a possible mode of combustion, however, from the pressure curve we can conclude that the massive star undergoing 
PT via detonation
gives rise to a comparatively small QS. 
The results from the pressure maximum, the physical velocities, we get a density range
where the PT is more likely. 
The stars outside this range are either not very likely to suffer PT or the PT in such stars cannot commence from the center of the star.

We must emphasize here that in this study we are not performing any dynamical calculations nor we are doing any micro-physical analysis. Our results are purely based on the 
hydrodynamic study of the initial and final EoSs. Although, the retracing nature is a global phenomenon, the quantitative value of the results dependent on the choice of 
EoS sets we choose.
The details of the micro-physics lie in the EoS, and we do not study them here.
Also, the quantitative region of detonation, deflagration and the value of maximum mass limit inferred by the restricted density range for PT depends on the choice of the 
EoS employed. 

The maximum of the downstream variables and the retracing of the CA curve is quite robust. It seems that this is a characteristic of the CA
and it could be a global property and not only related to astrophysical scenarios. Such nature of the CA can also be checked for ideal and realistic EoSs
which can be verified in high energy experiments. Also the present study gave rise to some interesting phenomena and results which should be supplemented with dynamic studies.
Our present effort is towards this regard.

\section{Acknowledgments}
The authors are grateful to the SERB, Govt. of India for monetary support in the form of Ramanujan Fellowship (SB/S2/RJN-061/2015) and Early Career Research Award (ECR/2016/000161). 
RM and MI would also like to thank IISER Bhopal for providing all the research and infrastructure facilities.

\end{document}